\documentclass[twocolumn,superscriptaddress,secnumarabic,aps,prl,nobibnotes,groupedaddress,numerical]{revtex4-1}

\usepackage{graphicx}
\usepackage{float}
\usepackage[squaren]{SIunits}
\usepackage{verbatim}
\usepackage{mathrsfs}
\usepackage{comment}
\usepackage{braket}
\usepackage{bbold}
\usepackage{amsmath}
\usepackage{hyperref}
\usepackage{pdfpages}
\usepackage[normalem]{ulem}
\usepackage{xcolor}
\usepackage{times}
\usepackage{amsmath}

\begin{document}

\title{Incompressible polaritons in a flat band}

\author{Matteo Biondi, Evert P. L. van Nieuwenburg, Gianni Blatter, Sebastian D. Huber, and Sebastian Schmidt}
\affiliation{Institute for Theoretical Physics, ETH Zurich, 8093 Z\"urich, Switzerland}

\begin{abstract}
We study the interplay of geometric frustration and interactions in a non-equilibrium photonic lattice system exhibiting a polariton flat band as described by a variant of the Jaynes-Cummings-Hubbard model. We show how to engineer strong photonic correlations in such a driven, dissipative system by quenching the kinetic energy through frustration. This produces an incompressible state of photons characterized
by short-ranged crystalline order with period doubling. The latter manifests itself in strong spatial correlations, i.e., on-site and nearest-neighbor anti-bunching combined with extended density-wave oscillations at larger distances. We propose a state-of-the-art circuit QED realization of our system, which is tunable in-situ.
\end{abstract}

\maketitle

Over the last decade there has been a surge of interest in realizing strongly correlated states of light in interacting photonic lattices for quantum simulations and the study of 
non-equilibrium many-body physics \cite{hartmann2006,greentree2006,angelakis2007,carusotto2009,gerace2009,hartmann2010,schmidt2010*2,nissen2012,sieberer2013,boite2013,jin2013,otterbach2013} (for two recent reviews, see \cite{houck2012,schmidt2013*2}).
Effective photon-photon interactions can be engineered in these systems utilizing strong light-matter couplings in various cavity/circuit QED platforms, e.g., with atoms \cite{birnbaum2005}, excitons \cite{reinhard2012,carusotto2013}  or superconducting qubits \cite{lang2011,hoffman2011}. Arranging cavities and atoms/qubits on a lattice offers the opportunity to engineer strongly correlated states of photons in various geometries with local control over coherent as well as dissipative dynamics. The driven dissipative nature of photonic systems then allows for direct experimental, non-invasive access to the complete density matrix, e.g., temporal and spatial correlation functions \cite{eichler2011*2}.

A particularly challenging and interesting problem of many-body physics concerns the study of frustrated lattices. Frustration refers to the impossibility of satisfying simultaneously all constraints implied by a Hamiltonian, which are imposed, e.g., by geometry, disorder or interactions. This typically leads to macroscopically degenerate ground-states, which are sensitive to small perturbations and thus define a challenging minimization problem. Conversely, frustration often gives rise to interesting strongly correlated phenomena and the emergence of fascinating non-trivial structures, e.g., in quantum magnetism \cite{Lieb1989_fb,Mielke1991_fb,Tasaki1992_fb,morris2009,nisoli2013}, quantum hall systems \cite{neupert2011,tang2011,petrescu2012}, Josephson junctions \cite{sigrist1995, Feigel'man_tetJJ_2004} or ultra-cold atoms \cite{wu2007,huber2010,Apaja2010_fb,Tovmasyan2013,zhu2014}. 
In this work, we make use of
geometric frustration to boost interactions and show that photons pumped into
the flat band of a photonic lattice form an incompressible state of light
with non-trivial spatial correlations at the onset of crystallization. This steady state cannot follow from energy minimization, but originates under non-equilibrium conditions with balanced drive and dissipation.

First realizations of interacting photonic lattices have recently been engineered based on superconductor as well as semiconductor technologies \cite{tanese2013,abbarchi2013,jacqmin2014,raftery2014,eichler2014*2}.
Motivated by these achievements, we study a 1D qubit-cavity chain, where qubits couple to photons in every other cavity. Such a Jaynes-Cummings-Hubbard (JCH) system \cite{greentree2006,angelakis2007,aichhorn2008,knap2010,hohenadler2011,hohenadler2012,schmidt2009,koch2009,schmidt2010,grujic2012,schmidt2013,Zhu2013} can be viewed as a quasi-1D cut through 
a 2D Lieb lattice \cite{Lieb1989_fb}, where the qubit represents one of the sites in the unit cell and simultaneously generates the frustration leading to the flat band as well as the photon-photon interaction. This setup is readily realizable with state-of-the-art circuit QED technology, where the lattice dispersion as well as the strength of the effective photon-photon interactions can be tuned in-situ by simply changing the qubit-resonator detuning \cite{schmidt2013*2}. 
We show that one of the Bloch bands of the array can be tuned from dispersive to completely dispersionless, i.e., flat in the entire Brillouin zone. This flat band arises due to destructive quantum interference and generates a macroscopic set of degenerate and localized plaquette states \cite{Mielke1991_fb,Tasaki1992_fb} (further information on creating flat bands can be found in the supplemental material (SM), which includes Refs.~\cite{Chalker2010_fb,Bergman2008_fb,Mielke_tasaki1993fb,Emery1987_hTc,Schollwoek_2011,zwolak_tebd_2004}). Similar flat bands were recently observed in non-interacting 2D laser and micro-pillar arrays \cite{guzman-silva2014,jacqmin2014}.

Here, we investigate the effects of strong photonic interactions in such a non-equilibrium geometrically-frustrated system by using projective methods as well as a time-evolving block decimation (iTEBD) algorithm \cite{Vidal2007,Orus_itebd_2008}. We find that geometric frustration strongly enhances photon repulsion on the lattice and pushes the system towards an incompressible state characterized by short-ranged crystalline order with period doubling. Incompressibility is signaled by the appearance of an extended plateau in the average polariton excitation number as a function of drive strength, whose height is determined solely by the geometry of the lattice. Crystallization manifests itself in strong spatial photonic correlations, i.e., on-site and nearest-neighbor anti-bunching combined with extended density-wave oscillations at larger distances. Interestingly, we find that the correlation length of these oscillations can be increased when decreasing the light-matter coupling strength $g$ with respect to the photon hopping rate $J$.

\begin{figure}[b]
\centering
\includegraphics[width=0.475\textwidth]{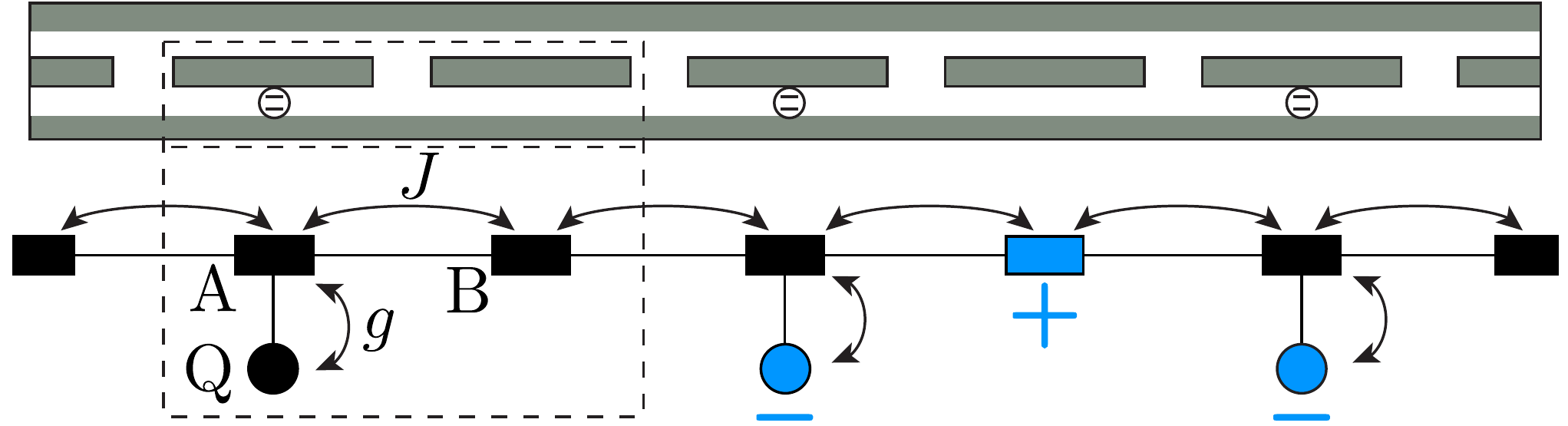}
\caption{(color online). Top: Sketch of a transmission line resonator chain including superconducting qubits in every other resonator. Bottom: Simplified lattice representation. Two resonators of type A and B (rectangles) are coupled by the photon hopping rate $J$. Qubits (circles) are only coupled to the A cavities with strength $g$. The dashed lines show one unit cell of the array and the blue symbols mark a localized plaquette state as discussed in the text ($\pm$ denote the corresponding phases in the wavefunction), see Eq.~\eqref{lambdastate}.\label{fig1}}
\end{figure}
\begin{figure}[b]
\centering
\includegraphics[width=0.4\textwidth]{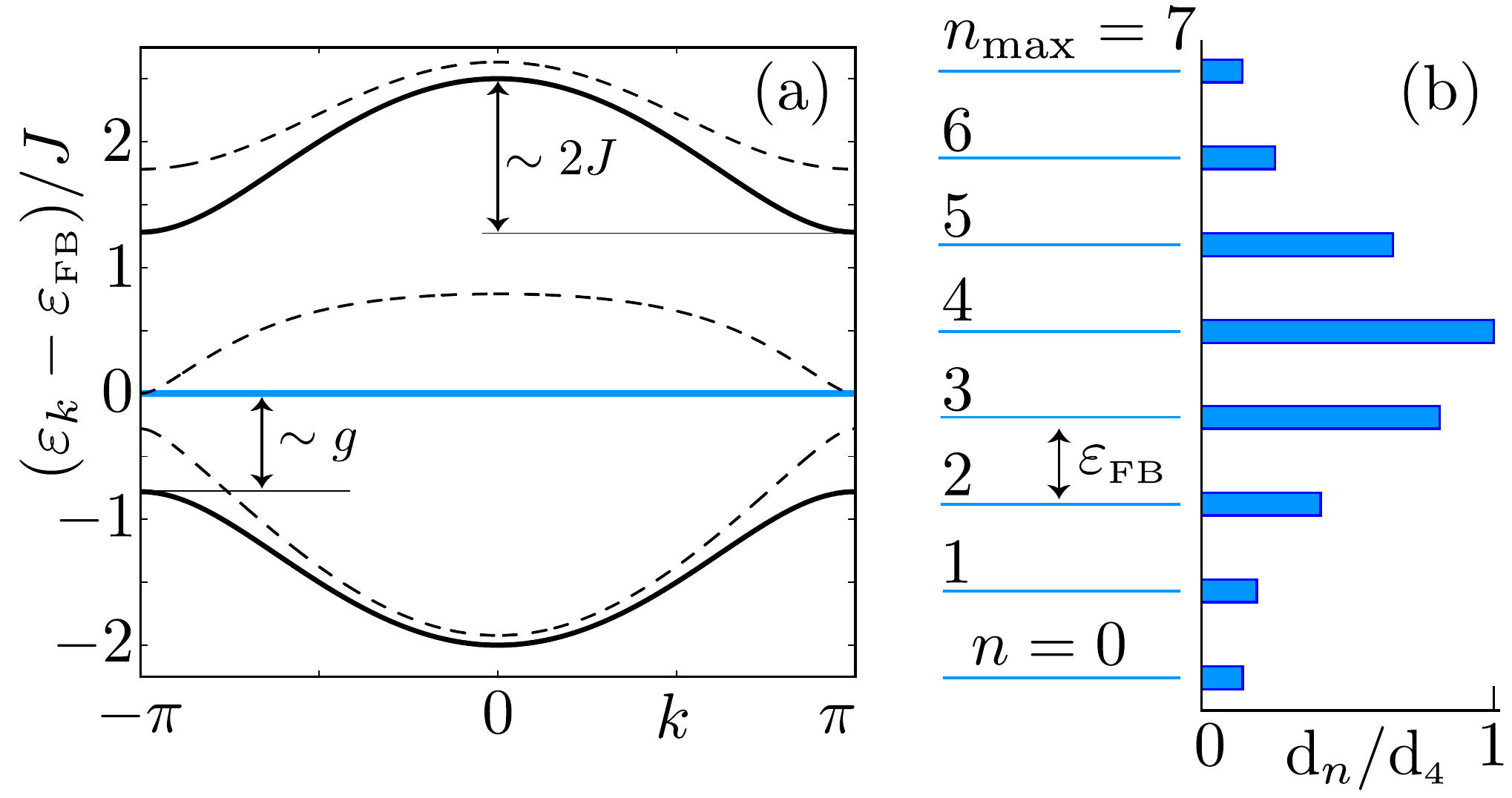} 
\caption{(color online). (a) Single particle dispersion of the lattice in Fig.~\ref{fig1}. For $\delta_{\rm \scriptscriptstyle QB} = J$ (dashed) all three bands are dispersive. For $\delta_{\rm \scriptscriptstyle QB}= 0$ (solid) the middle band (blue) becomes flat corresponding to a set of degenerate single-particle plaquette states. (b) Many-body eigenstates associated with the flat band are constructed from products of plaquettes (see main text). They form an equally spaced multi-level system indexed by the particle number $n=0,..,n_\text{max}$ with degeneracies $\text{d}_n = \binom{2n_\text{max}-n}{n}$ and largest filling $n_{\rm max}=(N+1)/2$. Here we show the case for $N=13$ unit cells.\label{fig2}}
\end{figure}
We study a variant of the driven dissipative JCH model, i.e.,
\begin{eqnarray}
\label{hammodel}
H = \sum_{j=1}^N \sum_{\alpha=\rm\scriptscriptstyle A,B} h_{j\alpha} + J \sum_{j=1}^{N-1} \left[(a_{j} + a_{j+1})b^\dagger_{j} + \text{H.c.}\right]
\end{eqnarray}
where $h_{j\alpha}$ denote the on-site Hamiltonians for resonators of type A and B with qubits at site Q coupling only to the A sites, i.e., $h_{j \rm \scriptscriptstyle A} = \Delta_{\rm \scriptscriptstyle A}a^\dagger_{j} a_{j} + \Delta_{\rm \scriptscriptstyle Q}\sigma^+_{j} \sigma^-_{j} + (ga^\dagger_{j} \sigma^-_{j} \! + f a_j + \text{H.c.})$ and $h_{j \rm \scriptscriptstyle B} = \Delta_{\rm \scriptscriptstyle B}b^\dagger_{j} b_{j} + f(b_j + \text{H.c.})$. The bosonic operators $a_{j}$ ($b_j$) annihilate a cavity photon at site A (B) in unit cell $j=1,\dots,N$. The second term in \eqref{hammodel} describes photon hopping between nearest neighbor resonators at a rate $J$. The qubits are represented by Pauli operators $\sigma^-_{j}$ and couple to the A photons with strength $g$. All cavities are subject to a coherent drive of strength $f$ described by the last terms in $h_{j\alpha}$. In a frame rotating with the drive frequency $\omega_{\rm \scriptscriptstyle D}$ the bare cavity and qubit frequencies $\omega_{\rm \scriptscriptstyle X}$ are renormalized to $\Delta_{\rm \scriptscriptstyle X} =  \omega_{\rm \scriptscriptstyle X} - \omega_{\rm \scriptscriptstyle D}$, with X = A, B, Q. Cavity dissipation is taken into account using a Lindblad master equation for the density matrix, i.e., $\dot{\rho} = -i[H,\rho] + \left(\kappa/2\right)\sum_{j}(\mathcal{D}[a_{j}]\rho + \mathcal{D}[b_{j}]\rho)$, with the Lindblad operator $\mathcal{D}[a]\rho = 2a\rho a^\dagger - a^\dagger a \rho - \rho a^\dagger a$ and the photon decay rate $\kappa$. Here we neglect spontaneous emission and dephasing of the qubits, which can be substantially suppressed with respect to cavity decay \cite{koch2007}. Fig. \ref{fig1} shows an implementation of our model based on state-of-the-art circuit QED technology \cite{schmidt2013*2}. A similar geometry can be realized using semiconductor micro-pillar arrays, where the qubit site is replaced with a nonlinear cavity (see the SM fur further information) \cite{carusotto2013,jacqmin2014}.

We start with the discussion of the single-particle spectrum of \eqref{hammodel} in the absence of drive ($f=0$) and dissipation ($\kappa=0$). For that purpose, we write a common Fourier transform $\boldsymbol{\psi}_{j} = (1/\sqrt{N})\sum_k e^{ikj} \boldsymbol{\psi}_{k}$, with $\boldsymbol{\psi}_{j} = [\,a_{j},\,b_{j},\,\sigma^-_{j}\,]^T$ and impose periodic boundary conditions to obtain the $k$-space representation of the lattice Hamiltonian, i.e., $H = \sum_{k} \boldsymbol{\psi}_{k}^\dagger h_k  \boldsymbol{\psi}_{k}$, with $k = 2n\pi/N$, $n = -N/2,\dots, N/2-1$. The eigenvalue equation for $h_k$ yields three bands, which are plotted in Fig.~\ref{fig2}(a). For the general case with $\delta_{\rm \scriptscriptstyle QB} = \omega_{\rm \scriptscriptstyle Q}-\omega_{\rm \scriptscriptstyle B}\neq 0$ all bands are dispersive (dashed lines). However, if $\delta_{\rm \scriptscriptstyle QB}=0$ the middle band turns flat with energy $\varepsilon_{\rm \scriptscriptstyle FB} = \omega_{\rm \scriptscriptstyle B}$ while the other two remain dispersive with energies $\varepsilon_{k}^\pm  = \omega_{\rm \scriptscriptstyle B} +\delta_{\rm \scriptscriptstyle AB}/2 \pm \sqrt{2J^2(1 + \cos k)+ g^2 + \delta_{\rm \scriptscriptstyle AB}^2/4}$, where $\delta_{\rm \scriptscriptstyle AB} = \omega_{\rm \scriptscriptstyle A} - \omega_{\rm \scriptscriptstyle B}$. The flat band eigenstates can be written as
\begin{equation}
\ket{\Lambda_{j}} = \frac{1}{\sqrt{g^2 + 2J^2}}\big[ g\,b_{j}^\dagger -J(\sigma_{j}^+ + \sigma_{j+1}^+)\big]\ket{\text{vac}},
\label{lambdastate}
\end{equation}
which describes a localized plaquette state defined by one B and two neighboring Q sites (see Fig. \ref{fig1}). The flat band arises due to the destructive interference between a photon hopping process from resonator B to A ($\sim J$) and the excitation transfer due to the coupling of qubit Q to the resonator A ($\sim g$). As a consequence the A cavities remain completely dark, such that an excitation originally localized at one end of the chain does not disperse and/or propagate to the other end.

In the following, we are interested in the interplay of frustration and interactions in the non-equilibrium steady state (NESS) of the system. In equilibrium, thermalization would lead to a zero temperature ground state not involving flat band
states as these do not reside at the lowest energy. In a non-equilibrium setup however, the coherent drive can excite
the flat band by keeping the drive frequency resonant with the flat band energy, i.e., $\omega_{\rm \scriptscriptstyle D} = \varepsilon_{\rm \scriptscriptstyle FB}$. In order to take into account states resonant with the drive, we construct from \eqref{lambdastate} the eigenstates of the Hamiltonian \eqref{hammodel} (for vanishing drive amplitude, i.e., $f=0$) with energies that are integer multiples of $\varepsilon_{\rm \scriptscriptstyle FB}$, and project the Lindblad equation on this subspace. Apart from the single-particle states in \eqref{lambdastate}, these states are products of non-overlapping plaquettes, e.g., the two particle states $\ket{\psi_2}\sim \ket{\Lambda_1}\ket{\Lambda_3},\ket{\Lambda_1}\ket{\Lambda_4}\dots$ with energy $2\varepsilon_{\rm \scriptscriptstyle FB}$, the three particle states $\ket{\psi_3}\sim \ket{\Lambda_1}\ket{\Lambda_3}\ket{\Lambda_5},\dots$ with energy $3\varepsilon_{\rm \scriptscriptstyle FB}$ etc. (for further details see the SM). 
The energetically highest lying state is the density-wave $\ket{\Psi_{\rm dw}} = \prod_{j=1}^{n_\text{max}}\ket{\Lambda_{2j-1}}$ with energy $\varepsilon_{\rm dw} = n_\text{max}\varepsilon_{\rm \scriptscriptstyle FB}$ and particle number $n_\text{max}=(N+1)/2$, i.e., filling per lattice site $\nu_{\rm dw} = n_\text{max}/(3N)=1/6 + \mathcal{O}(1/N)$. 
This special ladder of flat band states with degeneracies $\text{d}_n = \binom{2n_\text{max}-n}{n}$, where $n$ is the particle number of each state, is shown in Fig.~\ref{fig2}(b) for $N=13$ unit cells. All eigenstates with $n>n_\text{max}$ belong to dispersive bands and are gapped from the flat-band ladder due to the nonlinearity induced by the light-matter coupling $g$. 

\begin{figure}[t]
\centering
\includegraphics[width=0.4\textwidth]{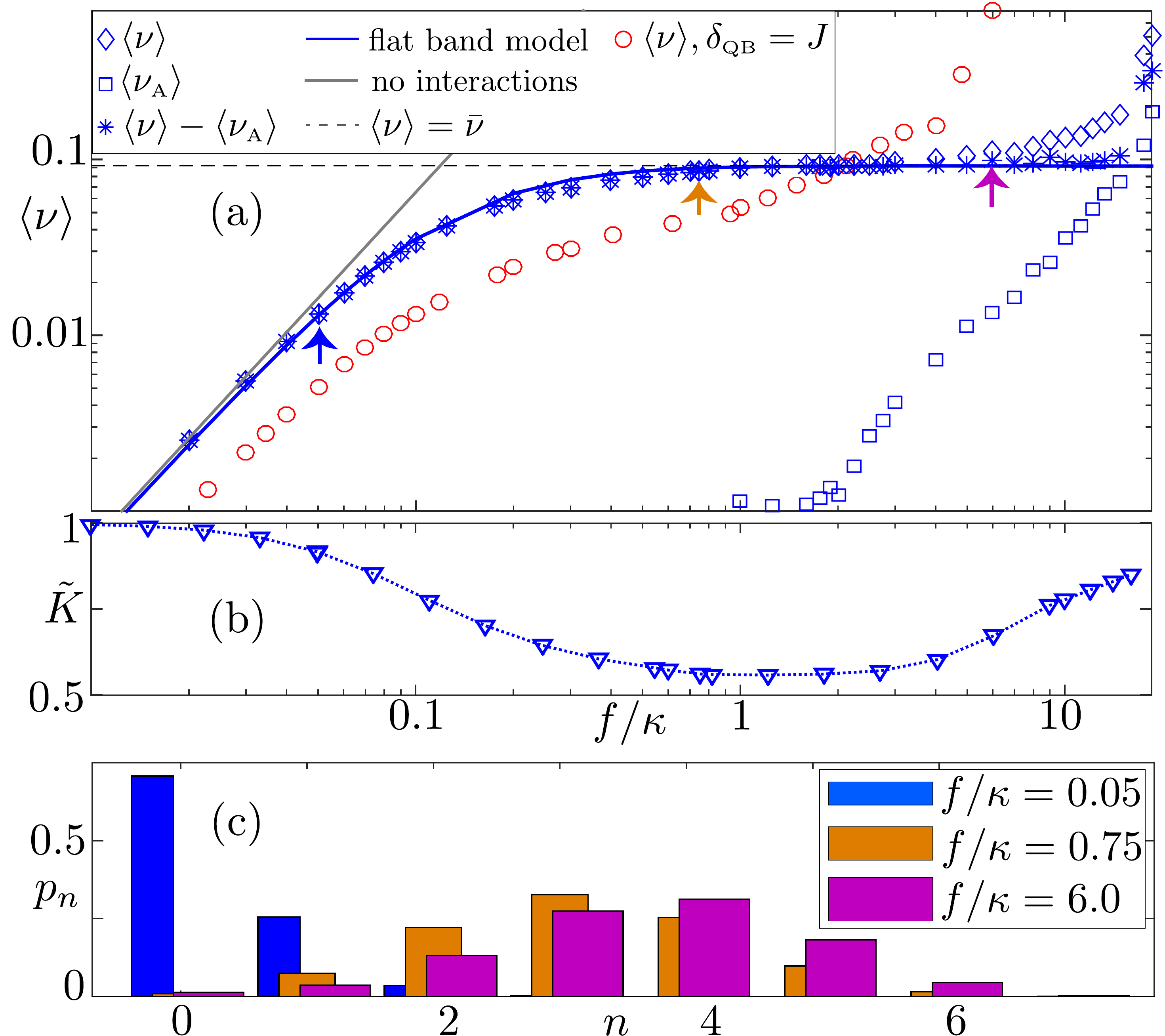}  
\caption{(color online). (a) Excitation number $\langle \nu \rangle = \sum_{\rm \scriptscriptstyle X}\langle\nu_{\rm \scriptscriptstyle X} \rangle$, with $\nu_{\rm \scriptscriptstyle X} = n_{\rm \scriptscriptstyle X}/(3N)$, $n_{\rm \scriptscriptstyle X} = \sum_{j} x_j^\dagger x_j$ (X = A, B, Q and $x_j = a_j,b_j,\sigma^-_j$) in the steady state as a function of pump strength $f/\kappa$. Shown are results obtained from projection of the density matrix on the flat band eigenspace for a system with $N=13$ unit cells and open boundary conditions (solid line) and from iTEBD simulations of the infinite system at zero detuning $\delta_{\rm \scriptscriptstyle QB}=0$ (blue symbols) and finite detuning $\delta_{\rm \scriptscriptstyle QB} = J$ (circles). The plateau at $\langle\nu\rangle\approx \bar{\nu}\approx 1/12$ is associated with a suppression of number fluctuations $\tilde{K} = [\langle (\sum_{\rm \scriptscriptstyle X} n_{\rm \scriptscriptstyle X})^2\rangle - (\sum_{\rm \scriptscriptstyle X} \langle n_{\rm \scriptscriptstyle X} \rangle)^2]/ \sum_{\rm \scriptscriptstyle X} \langle n_{\rm \scriptscriptstyle X} \rangle$ as shown in (b). The plateau is extended for the difference $\langle\nu\rangle - \langle\nu_{\rm \scriptscriptstyle A}\rangle$ (asterisks), but almost vanishes in the dispersive case (circles). (c) Probability $p_n$ of finding $n$ excitations in the lattice as calculated within the flat band model for the pump strengths marked with arrows in (a). Other parameters: $g/J=1$, $\delta_{\rm \scriptscriptstyle AB}/J=0.5$, $\kappa/J=0.05$. \label{fig3}}
\end{figure}

Due to the coherent drive with $\omega_{\rm \scriptscriptstyle D}=\varepsilon_{\rm \scriptscriptstyle FB}$ we expect states belonging to  the flat band ladder to mostly contribute to the NESS at small and intermediate drive strength. In Fig.~\ref{fig3} we show the average excitation number per lattice site $\langle \nu \rangle$ (for a formal definition see caption of Fig.~\ref{fig3}) as a function of pump strength $f/\kappa$. At weak pump $f\ll \kappa$, the results of the projected model (solid blue line) agree with the analytical expression $\langle \nu \rangle \approx (4f^2/\kappa^2)[\,1 + (4J^2 + \kappa^2/4)/g^2\,]$ (straight solid line), which is obtained from a perturbative calculation of the steady state to leading order in $f/\kappa$. At stronger pump, however, the system saturates at a filling $\langle\nu\rangle\approx \nu_{\rm dw}/2\approx 1/12$ resulting in an extended plateau centered around $f\sim \kappa$. This plateau can be interpreted as an incompressible state of photons with $\partial \langle \nu \rangle/\partial f \approx 0$, as we now explain in more detail.
The height of the plateau is largely independent of $g$ and $J$ and determined mostly by the geometry of the lattice. This can be understood by looking at the excitation number distribution $p_n$ of finding $n$ excitations in the lattice, shown in Fig.~\ref{fig3}(c). At weak pumping the distribution is peaked at low excitation numbers and shifts to larger $n$ for increasing pump strength. At strong pumping it saturates and resembles approximately the degeneracies $\text{d}_n$ shown in Fig.~\ref{fig2}(b), i.e., all states are almost
equally occupied similar to a two-level system saturating half way between ground and excited state \cite{bishop2009}. The saturated average excitation number is thus calculated as $\bar{n} \approx (\sum_{n=0}^{n_{\text{max}}}n\,\text{d}_n)/(\sum_{n=0}^{n_{\text{max}}}\text{d}_n)= (1-1/\sqrt{5})(N/2)$ corresponding to roughly half the density-wave filling $\bar{\nu}=\bar{n}/(3N)\approx \nu_{\rm dw}/2$ (horizontal dashed line in Fig.~\ref{fig3}(a)). The incompressible state thus originates from an unconventional photon blockade on a frustrated lattice arising from a saturation of the flat band ladder shown in Fig.~\ref{fig2}(b).

\begin{figure}[t]
\centering
\includegraphics[width=0.4\textwidth]{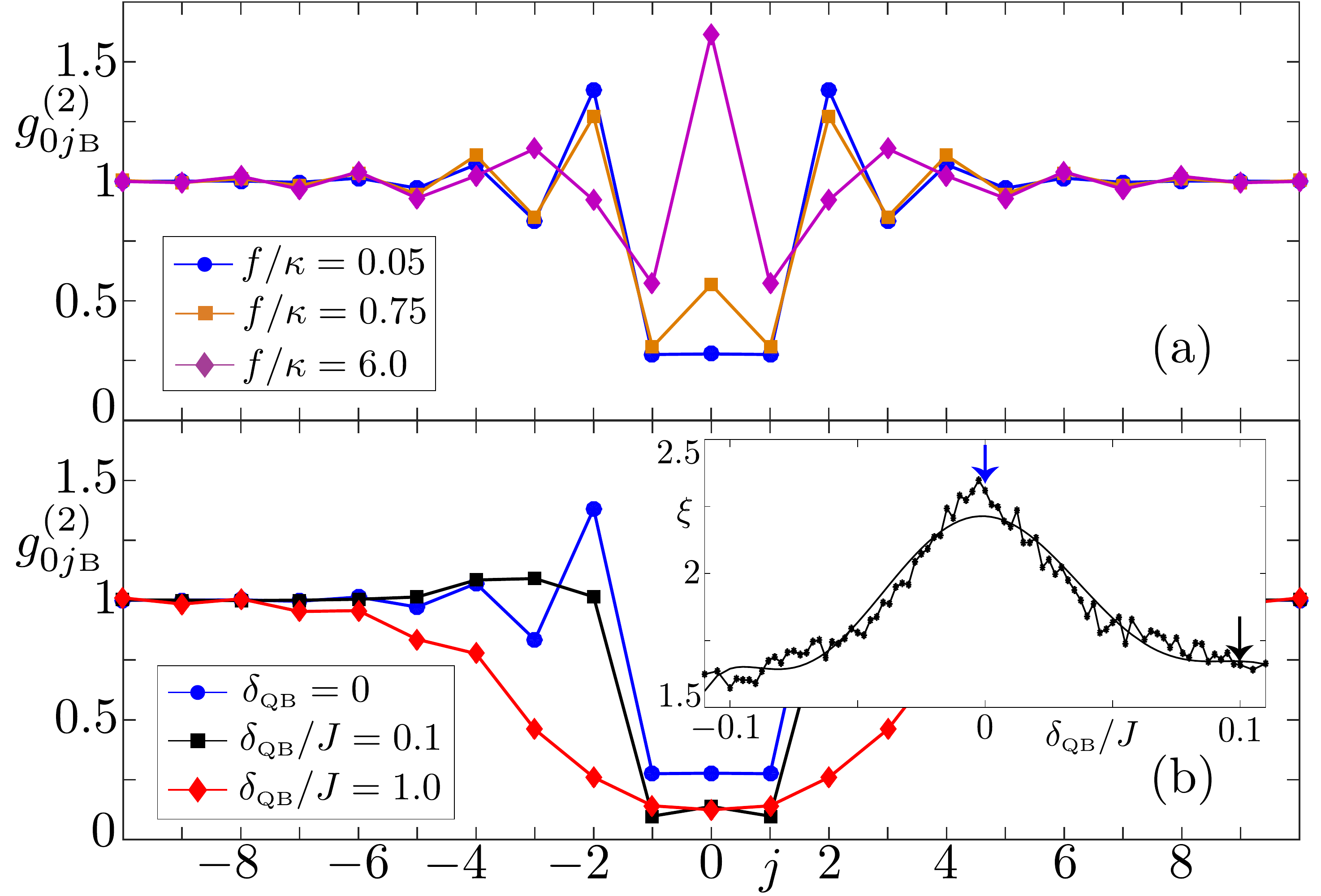}  
\caption{(color online). Correlation function of photons emitted by the B sites $g^{(2)}_{0j\rm \scriptscriptstyle B} = \langle b^\dagger_0 b^\dagger_j b_0 b_j\rangle/\langle b^\dagger_0b_0\rangle\langle b^\dagger_j b_j \rangle$ for different drive strength's $f/\kappa$ at fixed detuning $\delta_{\rm \scriptscriptstyle QB} =0$ (upper panel) and different detunings $\delta_{\rm \scriptscriptstyle QB}/J$ at fixed drive strength $f/\kappa=0.05$ (lower panel) as calculated with iTEBD. The density-wave oscillations correspond to a period doubling with respect to the unit cell of the underlying lattice. In (b) the drive stays resonant with the top of the middle band. The inset shows the length of the density-wave oscillations $\xi$ obtained from an exponential fit. Arrows mark the corresponding values in the main figure. Other parameters chosen as in Fig.~\ref{fig3}.
\label{fig4}}
\end{figure}

We confirm this picture by numerical simulations employing an open system version of the iTEBD algorithm \cite{Vidal2007,Orus_itebd_2008} (for technical details see SM). In Fig.~\ref{fig3}(a) the projected model agrees with the exact numerics (diamonds) well into the plateau, thus verifying the incompressible state of photons, where fluctuations of the excitation number are reduced (see Fig.~\ref{fig3}(b)). For an even stronger pump ($f\gg \kappa$), the dispersive bands start to contribute to the NESS leading to a destruction of the incompressible state. This is also signaled by an increasing occupation of the A cavities (squares). In this regime, the projected model becomes invalid and the full numerics very costly as the local Hilbert space cutoff needs to be increased substantially. The interesting details of this crossover are subject of future work. 

We now investigate the spatial order of the steady state by studying the second-order coherence function (density-density correlator) of the B sites, i.e., $g^{(2)}_{ij\rm \scriptscriptstyle B} = \langle b^\dagger_i b^\dagger_j b_i b_j\rangle/\langle b^\dagger_ib_i\rangle\langle b^\dagger_j b_j \rangle$. Fig.~\ref{fig4} shows the spatial correlations of the central B site ($i=0$) with its neighbors as calculated with iTEBD. At weak and intermediate pump strength $f/\kappa$ we find local ($j=0$) as well as nearest neighbor ($j=\pm1$) anti-bunching, which represent a signature of photon blockade and incompressibility, i.e., the resistance of the system to accept simultaneously two pump photons entering the chain either on the same or on neighboring plaquettes (which share a qubit, see Fig.~\ref{fig1}). Thus, if a photon is present at a B site of the chain, every other B site is less occupied due to effective photon-photon interactions resulting in polaritonic density-wave like order. At larger distances, density-wave order manifests itself in correlations alternating between bunching ($g^{(2)}_{0(2j)\rm \scriptscriptstyle B} \!> 1$) and anti-bunching ($g^{(2)}_{0(2j+1)\rm \scriptscriptstyle B}\!< 1$) with a period doubling of two unit cells, leading to an incipient crystalline state of light. This can be interpreted as the non-equilibrium counterpart of a charge density wave appearing in the ground-state of an electronic or atomic system with a flat lowest-energy band, e.g., in a sawtooth or Kagome lattice \cite{huber2010}. 

\begin{figure}[t]
\centering
\includegraphics[width=0.4\textwidth]{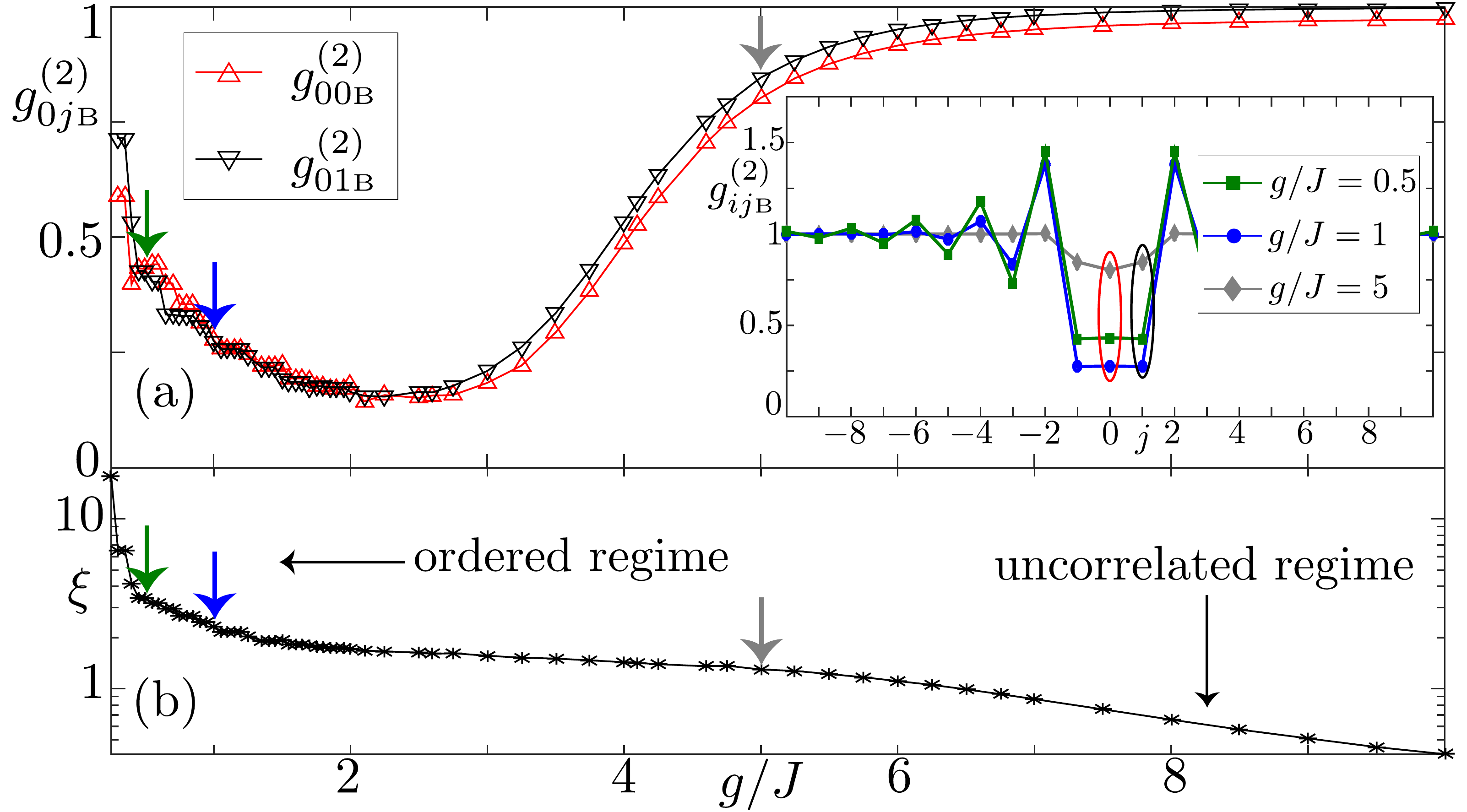}  
\caption{(color online). (a) On-site and nearest-neighbor correlator $g^{(2)}_{00\rm \scriptscriptstyle B}, g^{(2)}_{01\rm \scriptscriptstyle B}$ as a function of $g/J$ for $f/\kappa = 0.05$ (blue arrow in Fig.~\ref{fig3}). The inset shows the complete spatial dependence of the coherence function for the $g/J$ values marked by arrows in the main figure. (b) Correlation length as a function of $g/J$ obtained from an exponential fit of the correlator. Other parameters chosen as in Fig.~\ref{fig3}.
\label{fig5}}
\end{figure}

In a regular one-dimensional Jaynes-Cummings array interactions vanish when $g\ll J$ \cite{nissen2012}. Interestingly, for the flat band the converse is true, as the ratio $g/J$ determines the polaritonic nature of the plaquette states, which are qubit-like and thus strongly interacting when $g\ll J$ (see Eq.~\eqref{lambdastate}). This remarkable effect determines the spatial extent of the density-wave correlations. As shown in Fig.~\ref{fig5}, the correlation length shrinks when the flat band becomes photon-like ($g\gg J$), while it grows steeply in the opposite limit ($g\ll J$). On the technical level, the projection on the flat band (see SM) modifies the drive strength according to $f/\kappa\rightarrow (f/\kappa)\sqrt{1 + 2J^2/g^2}$, thus effectively increasing the drive strength when $g/J$ decreases. This entails a larger contribution of the high-energy density-wave state $\ket{\Psi_{\rm dw}}$ (with infinite correlation length) to the NESS. At the same time the gap to the other bands closes as $g/J\rightarrow0$ leading to a destruction of the photon blockade. Consequently, we find the strongest anti-bunching for the fully mixed polaritonic case when $g \sim \sqrt{2}J$.

Finally, we show that the signatures of geometric frustration, incompressibility and crystalline order vanish when the flat band becomes dispersive, i.e., when $\delta_{\rm \scriptscriptstyle QB} \neq 0$ (the drive stays resonant with the top of the band). For $\delta_{\rm \scriptscriptstyle QB} =J$ (compare with the dashed lines in Fig.~\ref{fig2}(a)), we observe in Fig.~\ref{fig3}(a) and \ref{fig4}(b) that the plateau as well as density-wave like correlations completely disappear. The latter are replaced by a broad and rather featureless anti-bunching dip in Fig.~\ref{fig4}(b). Indeed, the correlation length of the density-wave oscillations drops quickly from its maximum flat band value to roughly one unit cell (see inset of Fig.~\ref{fig4}).

In summary, we have shown that geometric frustration in a photonic lattice pushes the system towards an incompressible state of light characterized by short-ranged crystalline order with period doubling. We have proposed the simplest model of a frustrated quasi-1D lattice based on a circuit QED architecture realizable with state-of-the-art technology and easily extensible to two dimensions, e.g., to study topological effects. A variant suitable for a realization of our proposal based on semiconductor micro-pillar arrays \cite{carusotto2013,jacqmin2014} is described in the supplemental material and has recently been realized experimentally \cite{baboux2015}. The onset of long-range correlations motivates another interesting question for future work, i.e., whether super-solid behavior of light (coexistence of superfluidity and density-wave order) could be observed in a flat band without the need of explicitly engineering nearest-neighbor interaction terms in the Hamiltonian \cite{otterlo1995,jin2013}. Our proposal thus paves the way for quantum simulations \cite{georgescu2014} of frustrated systems far from equilibrium and the realization of strongly correlated, exotic states of light with non-trivial spatial correlations.

\begin{acknowledgments}
We acknowledge fruitful discussions with A. Amo, F. Baboux, J. Bloch, M. Bordyuh, C. Ciuti, K. LeHur, H. E. T\"ureci and G. Zhu and support from the Swiss NSF through an Ambizione Fellowship (SS) under Grant No.\ PP00P2-123519/1 and the NCCR QSIT (MB).
\end{acknowledgments}

\bibliography{fb_pol}

\end{document}